\documentclass[fleqn,usenatbib]{mnras}

\usepackage{newtxtext,newtxmath}

\usepackage[T1]{fontenc}

\DeclareRobustCommand{\VAN}[3]{#2}
\let\VANthebibliography\thebibliography
\def\thebibliography{\DeclareRobustCommand{\VAN}[3]{##3}\VANthebibliography}

\usepackage{graphicx}	
\usepackage{amsmath}	

\pdfoutput=1
\usepackage[T1]{fontenc}
\usepackage{graphicx}
\usepackage{wasysym}
\usepackage{esint}
\usepackage[authoryear]{natbib}
\usepackage[utf8]{inputenc}
\usepackage{threeparttable}
\usepackage{hyperref}
\usepackage{xspace}
\usepackage{url}
\usepackage{babel}
\usepackage{ulem}
\usepackage{newtxtext,newtxmath}
\usepackage{hyperref}	
\usepackage{mathtools}
\usepackage{xcolor}
\usepackage{multirow}
\usepackage{orcidlink}

\hypersetup{colorlinks=true,linkcolor=blue,citecolor=blue,filecolor=blue,urlcolor=blue,}

\newcommand{\auriga}{\textsc{Auriga}\xspace}
\newcommand{\arepo}{\textsc{Arepo}\xspace}

\newcommand{\subfind}{\textsc{subfind}\xspace}

\newcommand{\msun}{\mathrm{M}_\odot}
\newcommand{\bs}[1]{\boldsymbol{#1}}

\title[magnetic fields in satellite galaxies]{Environment matters: stronger magnetic fields in satellite galaxies}
\author[M. Werhahn et al.]{Maria Werhahn$^{1,}$\thanks{E-mail:
mwerhahn@mpa-garching.mpg.de}\orcidlink{0000-0003-4984-4389}, 
R\"udiger Pakmor$^1$%
\orcidlink{0000-0003-3308-2420},
Rebekka Bieri$^2$%
\orcidlink{0000-0002-4554-4488},
Freeke van~de~Voort$^3$%
\orcidlink{0000-0002-6301-638X},
Rosie Y. Talbot$^1$%
\orcidlink{0000-0001-9393-7879},
\newauthor Volker Springel$^1$%
\orcidlink{0000-0001-5976-4599}
\\
\\%
$^{1}$Max-Planck-Institut f\"ur Astrophysik (MPA), Karl-Schwarzschild-Str. 1, 85748 Garching, Germany\\%
$^{2}$Department of Astrophysics, University of Zurich, 8057 Zurich, Switzerland\\%
$^{3}$Cardiff Hub for Astrophysics Research and Technology, School of Physics and Astronomy, Cardiff University, Queen’s Buildings, Cardiff CF24 3AA, UK
}

\date{Accepted XXX. Received YYY; in original form ZZZ}

\pubyear{2025}

\begin{document}
\label{firstpage}
\pagerange{\pageref{firstpage}--\pageref{lastpage}}
\maketitle

\begin{abstract} 
Magnetic fields are ubiquitous in the universe and an important component of the interstellar medium. It is crucial to accurately model and understand their properties in different environments and across all mass ranges of galaxies to interpret observables related to magnetic fields correctly. However, the assessment of the role of magnetic fields in galaxy evolution is often hampered by limited numerical resolution in cosmological simulations, in particular for satellite galaxies.
To this end, we study the magnetic fields in high-resolution cosmological zoom simulations of disk galaxies (with $M_{200}\approx10^{10}$ to $10^{13}\,\msun$) and their satellites within the Auriga galaxy formation model including cosmic rays.
We find significantly higher magnetic field strengths in satellite galaxies compared to isolated dwarfs with a similar mass or star-formation rate, in particular after they had their first close encounter with their host galaxy.
These are stronger on average by factors of 2--8 when compared at the same total mass, with a large scatter, ranging up to factors of $\sim$15.
While this result is ubiquitous and independent of resolution in the satellites that are past their first infall, there seems to be a wide range of amplification mechanisms acting together.
Our result highlights the importance of considering the environment of dwarf galaxies when interpreting their magnetic field properties as well as related observables such as their gamma-ray and radio emission, the latter being particularly relevant for future observations such as with the SKA observatory.
\end{abstract}




\section{Introduction}

Magnetic fields are considered to be a dynamically relevant component of spiral galaxies, exhibiting a typical strength of a few $\mu$G and thereby having energy densities comparable to other energetically important components of the interstellar medium \citep[ISM;][]{2015Beck, 2023BrandenburgNtormousi}.
Their related observables, in particular synchrotron emission and Faraday rotation, allow us to estimate their strength and structure in the Milky Way (MW) and nearby star-forming galaxies.
Magnetohydrodynamical (MHD) simulations of galaxies play a major role in understanding the origin of magnetic fields and their role in the formation and evolution of galaxies \citep{2024Korpi-Lagg}. The amplification mechanism of magnetic fields has been studied both in isolated \citep[e.g.][]{2016Rieder,Butsky2017,2022Pfrommer} and cosmological set-ups \citep[e.g.][]{Pakmor2014,Pakmor2017,2017Su, Martin-Alvarez2018, 2024Pakmor}.
In addition, simulations show that they have a significant impact on remnants of mergers provided the resolution is sufficiently high \citep{2021Whittingham}, as well as on the physical properties of the circumgalactic medium (CGM) of galaxies \citep{2021VanDeVoort}.

Another important non-thermal constituent of the ISM are cosmic rays (CRs), which have a comparable energy density in the mid-plane of the MW to the thermal, turbulent and magnetic components \citep{1990BoularesCox,2013Zweibel}. 
CRs interact through wave-particle interactions with the ISM, leading to an exchange of energy and momentum with it. Therefore, they can have a significant effect on galactic dynamics by driving winds \citep[see e.g.][for recent reviews]{2017Zweibel, 2023RuszkowskiPfrommer_Review}.
Their interaction with the gas and magnetic fields of the ISM induce the emission of non-thermal radiation, ranging from the radio to the gamma-ray regime. Observations of this non-thermal radiation can potentially be used to constrain the CR content and how they interact with the ISM, thereby allowing us to assess their importance as a feedback process in star-forming galaxies.

This has been studied, in particular, within CR-MHD simulations that simultaneously account for non-thermal radiation processes and their observational constraints \citep[e.g.][]{2019Chan, 2020Buck,2021WerhahnII, 2021WerhahnIII, 2022Pfrommer, 2024Ponnada, 2024ChiuSandy}.
To model CR transport and the related emission, in particular from CR electrons and positrons, it is crucial to understand magnetic fields, their structure and amplification mechanisms.
Observations of an excess in radio emission in ram-pressure stripped galaxies in galaxy clusters \citep[e.g.][]{1991Gavazzi,1999Gavazzi,2009Murphy,2013Vollmer,2020Chen,2022aIgnesti,2024Edler} suggest that the environment can significantly affect the observed radio emission. We therefore aim to assess the impact of environment on galactic magnetic fields, which could in turn influence the emitted radio synchrotron emission. In addition, it could also indirectly impact gamma-ray emission from CR protons due to the dependence of CR transport and Alfv\'en wave cooling on magnetic fields.
To this end, large cosmological boxes would be ideally placed to study this statistically, across a wide variety of galactic environments. However, this approach would lack the required resolution to study the dynamo processes that are likely the main driver to amplify small seed fields to the observed galactic values \citep{2022Martin-Alvarez, 2024Pakmor}. Therefore, we study cosmological zoom-in simulations of individual haloes in this paper, including CRs, where we also re-assess the resolution requirements for converged magnetic fields in isolated dwarf galaxies versus satellites.

The structure of this paper is the following. We first describe our simulation set-up in Section~\ref{Sec:Simulations}. Section~\ref{sec:B-fields in dwarfs} gives an overview of the different environment of satellites compared to isolated dwarfs and quantifies the difference in magnetic field strengths as a function of total mass and star formation rate (SFR). In Section~\ref{sec:B-field-evolution-turbulent-driving} we discuss the temporal evolution of the magnetic energy, quantify the turbulent driving scale via the second-order velocity structure function, and investigate the effect of resolution on the amplification. Finally, we discuss potential amplification mechanisms and the saturation strength of the magnetic field in Section~\ref{sec:discussion}, before we summarize our results in Section~\ref{sec:conclusion}. 
The numerical convergence of the magnetic fields in our simulations with and without CRs is discussed in Appendix~\ref{app:convergence-bfield}.
Additionally, we complement our analysis by magnetic and kinetic power spectra in Appendix~\ref{app:powerspectra}, and show radial profiles of the magnetic energy density in Appendix~\ref{app:radial_profiles}.

\section{Simulations}
\label{Sec:Simulations}

\begin{table}
\begin{tabular}{ |c|c|c|c| } 
\hline
Name & $M_{200}$ & Number of satellites & Mass range of satellites \\
 & $[\mathrm{M_\odot}]$ & &  $M_\mathrm{tot}\,[\mathrm{M_\odot}]$ \\
\hline
1e13-h7	 & $10^{13.11}$ & 33 &	$10^{9.39}$ to $10^{11.77}$ \\
1e13-h8  & $10^{12.99}$ & 18 &	$10^{9.40}$ to $10^{11.58}$\\ 
1e13-h3  & $10^{12.51}$ & 9 &	$10^{9.51}$ to $10^{10.53}$ \\
1e12-h5  & $10^{12.08}$ &  1 &  $10^{10.30}$ \\
1e12-h12 & $10^{12.04}$ &  4 &	$10^{9.63}$ to $10^{10.15}$ \\
1e11-h4  & $10^{11.44}$ &  2 &	$10^{8.90}$ to $10^{9.89}$ \\
1e11-h5  & $10^{11.46}$ & -  & - \\
1e11-h11 & $10^{10.98}$ & 1  &  $10^{10.30}$ \\
1e11-h10 & $10^{10.91}$ & -  & - \\
1e10-h9  & $10^{10.54}$ & -  & - \\
1e10-h11 & $10^{10.34}$ & -  & - \\
1e10-h8  & $10^{10.08}$ & -  & - \\
1e10-h12 & $10^{9.84}$  & -  & - \\
\hline
\end{tabular}
\caption{Overview of the simulated haloes and their star-forming satellites: name of the halo, $M_{200}$, number of star-forming satellites, and the range of their total masses at $z=0$.}
\label{table1}
\end{table}

Our cosmological zoom-in simulations are based on 13 haloes selected from the dark matter only EAGLE box \citep{2015Schaye} that are re-simulated with a much higher resolution. We use two sets of simulations: the one described in \citet{2024Pakmor}\footnote{Note that in comparison to their work, we exclude in this study halo 1e13-h4 due to a minor contamination by lower resolution dark matter particles.} and a second set of simulations of the same haloes that additionally include CRs (Bieri et al. in preparation). Both sets of simulations are run with the moving-mesh code \arepo \citep{2010Springel,2016aPakmor}, which adopts a second-order finite-volume scheme to solve the ideal MHD equations \citep{2011Pakmor,2013Pakmor}, using the \citet{1999Powell} approach for divergence control. The initial conditions include an initial comoving magnetic field strength of $10^{-14}\,\mathrm{G}$. The properties of the final galactic magnetic fields have, however, been shown to be largely insensitive to the seeding mechanism, the initial value, and the direction of the field \citep{Pakmor2014,2015Marinacci,2021Garaldi}.

Furthermore, our simulations employ the \auriga galaxy formation model \citep{2017Grand}. This includes an effective model for galactic winds as well as a framework accounting for feedback from active galactic nuclei, where black holes are seeded with a mass of $10^5\,h^{-1}\,\mathrm{M_\odot}$ in haloes with masses exceeding $5\times 10^{10}\,h^{-1}\,\mathrm{M_\odot}$ \citep[see ][ for details]{2017Grand}. We adopt a two phase treatment of the ISM with an effective equation of state \citep{2003SpringelHernquist}, including a stochastic model for star formation. Once a new star particle is created, we inject a fraction $\zeta_\mathrm{SN}$ of the supernova (SN) energy of $10^{51}$~ergs into CRs. We ran all haloes with a fiducial value of $\zeta_\mathrm{SN}=0.10$ and additionally re-ran six of them with $\zeta_\mathrm{SN}=0.05$.
CRs are modelled as a relativistic fluid with an adiabatic index of $4/3$. We further assume that their transport can be described by a combination of advection with the gas and diffusion relative to it in an anisotropic manner along magnetic field lines \citep{2017aPfrommer}, where we adopt a diffusion coefficient of $\kappa=10^{28}\,\mathrm{cm^2\,s^{-1}}$. We vary $\kappa$ for three haloes to a higher value of $3\times10^{28}\,\mathrm{cm^2\,s^{-1}}$ (for 1e10-h8, 1e11-h10, and 1e12-h12, see Table~\ref{table1}).
More realistically, in media without efficient damping of Alfv\'en waves, CR transport is mediated by interactions with Alfv\'en waves, limiting the CR transport speed to the Alfv\'en velocity, which would require more sophisticated modelling like in a two-moment description \citep[e.g.\ ][]{2018JiangOh,2019ThomasPfrommer, 2020Hopkins}. To emulate those losses due to interactions with Alfv\'en waves, we include an Alfv\'en loss term \citep[as derived in ][]{Wiener2013} to the CR energy equation \citep[see also ][Bieri et al.\ in preparation]{2010Sharma, 2012Uhlig, 2017Wiener, 2019Dubois, 2020Hopkins, 2020Buck, 2021Hopkins_SelfConf}.

We simulated 13 haloes with masses ranging from $M_{200}=10^{10}\,\msun$ to $10^{13}\,\msun$ at $z=0$. The halo mass $M_{200}$ refers to the mass within a sphere with radius $R_{200}$ which encompasses a region with an average density equal to 200 times the critical density of the universe.
Both for the simulations with and without CRs, we ran the dwarf galaxies with $M_{200}\leq 10^{10.6}\,\msun$ with a gas mass resolution of $8\times 10^2\,\msun$ (`level~2'), the medium mass haloes $ 10^{10.6}\,\msun \leq M_{200}\leq 10^{11.5}\,\msun$ with $6\times 10^3\,\msun$ (`level~3'), and the massive haloes with $M_{200}\geq 10^{12}\,\msun$ at a gas mass resolution of $5\times 10^4\,\msun$ (`level~4'). In addition, to test for convergence, we computed four haloes with CRs also at a lower resolution (see Fig.~\ref{fig:B_resolution}), and the halo 1e12-h12 at a higher resolution of $6\times 10^3\,\msun$. We ran all haloes except 1e12-h5 without CRs at a lower resolution level as well.
In the following, we will show our main results for the CR-MHD runs, which we will refer to as the `fiducial' simulations.

Subhaloes within a halo are identified as gravitationally bound particles using the \subfind algorithm \citep{2001Springel}. The most massive galaxy of the halo is the `central' galaxy, we refer to all other subhaloes as `satellites'.
We analyse all satellite galaxies which are star forming at $z=0$, i.e.\ which have a non-zero average SFR within the last 100~Myr. Our central galaxies exhibit 1--33 satellites (see Table~\ref{table1} for their host halo masses and total satellite masses, i.e. the sum of gas, stars and dark matter bound to the subhalo).

\section{Magnetic fields of satellites versus central galaxies}\label{sec:B-fields in dwarfs}

\begin{figure*}
    \centering
    \includegraphics{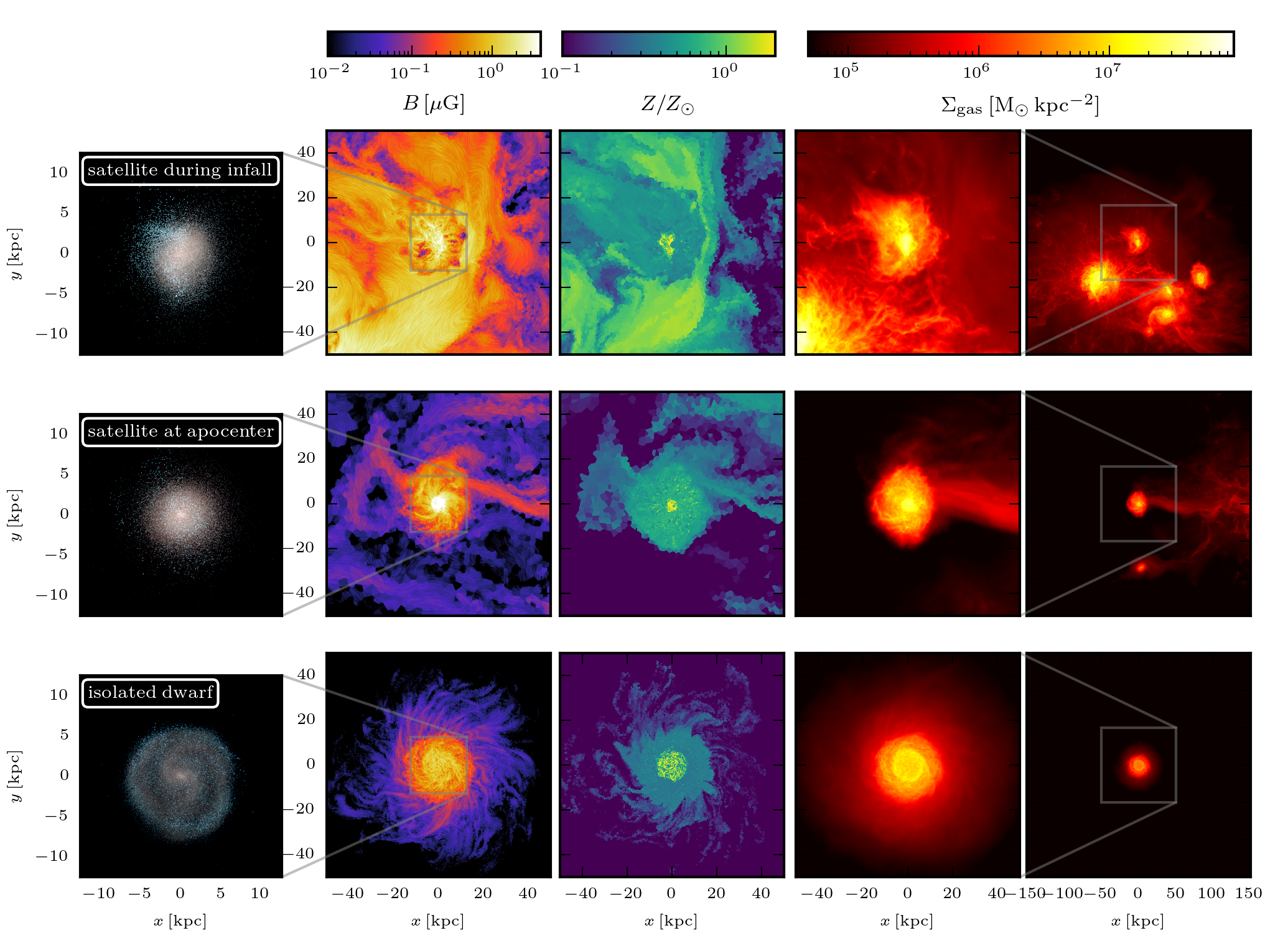}
    \caption{Maps of an example satellite at the time of its infall into the host halo (first row; $z=0.39,\ t_\mathrm{look}=4.31$~Gyr), at the apocentre of its orbit (second row; $z=0.14,\ t_\mathrm{look}=1.85$~Gyr), and an isolated dwarf with a similar stellar mass (third row; $z=0$). We show, from left to right, the stellar light projection (the red, green and blue colours in the stellar light projections represent the {\it K}-, {\it B}- and {\it U}-band luminosity of stars, respectively), a thin projection (of thickness 3~kpc) of the magnetic field strength and a slice of the metallicity, as well as the projected gas surface density in two different sized boxes. The orientation of the magnetic field in the second column is indicated by a relief created by the line integral convolution method \citep{1993CabralLeedom}. The first row shows maps rotated such that the satellite moves in the positive $x$-direction, while the second and third row show the galaxies face-on. The satellite exhibits a stronger magnetic field at apocentre than the isolated dwarf.
    }
    \label{fig:map}
\end{figure*}

Since we aim at analysing the effect of environment on the magnetic field properties of dwarf galaxies in this work, we compare our isolated dwarfs\footnote{The term `isolated dwarfs' refers to central galaxies of our low-mass haloes without any star-forming satellites, and within the same mass range as the satellites (see Table~\ref{table1}).} to satellite galaxies at different stages of their infall into the host halo's potential. In particular, we analyse the satellites during first approach vs. satellites after their first infall. After first infall means that the satellite has passed the point of its first minimum distance to the halo for at least 300~Myr. 

We first aim to gain a qualitative impression of the different environments that satellites encounter over time and compare it to an isolated dwarf. To this end, we present in Fig.~\ref{fig:map} maps of the satellite with the largest stellar mass at $z=0$ , i.e.\ the sum of all star particles that are bound to the subhalo ($M_\star=10^{9.05}\,\mathrm{M_\odot}$) of the high-resolution simulation (level~3) of halo 1e12-h12. We depict the satellite both during its first infall into the potential of its host halo ($z=0.39,\ t_\mathrm{look}=4.31$~Gyr, $M_\star=10^{8.92}\,\mathrm{M_\odot}$, $M_\mathrm{gas}=10^{9.58}\,\mathrm{M_\odot}$, $\dot{M}_\star=0.30\,\mathrm{M_\odot/yr}$) and at the apocentre of its orbit ($z=0.14,\ t_\mathrm{look}=1.85$~Gyr, $M_\star=10^{8.99}\,\mathrm{M_\odot}$, $M_\mathrm{gas}=10^{9.53}\,\mathrm{M_\odot}$, $\dot{M}_\star=0.13\,\mathrm{M_\odot/yr}$). 
The latter refers to the time of the first local maximum in the distance to the central galaxy as a function of time amongst the output snapshots.
In addition, we show an isolated dwarf (halo 1e10-h9) with $M_\star=10^{8.67}\,\mathrm{M_\odot}$, $M_\mathrm{gas}=10^{9.68}\,\mathrm{M_\odot}$ and $\dot{M}_\star=0.06\,\mathrm{M_\odot/yr}$ at $z=0$.

The gas surface density maps in the two right-hand columns of Fig.~\ref{fig:map} reveal the vastly different environments that the satellite encounters during infall. First, it comes close to the central galaxy of the halo (a MW-like galaxy) and also other satellites, whereas at its apocentre it almost appears to be a dwarf galaxy in isolation. 
However, the stellar light projections (left column of Fig.~\ref{fig:map}) are more centrally concentrated than the stellar light of the isolated dwarf galaxy with a similar stellar mass, which is shown in the third row of Fig.~\ref{fig:map}.
The magnetic field map during infall hints towards magnetic draping in front of the infalling satellite (the satellite is moving in the positive $x$-direction in this figure). After its first approach, the satellite exhibits a higher field strength than during the first infall. In particular, within a sphere of radius 5~kpc (10~kpc) around the centre of the satellite, the average magnetic field strength increases from 1.6 (0.8)~$\mu\mathrm{G}$ at $t_\mathrm{look}=5$~Gyr, to 1.9 (1.2) at $t_\mathrm{look}=4.31$~Gyr (as shown in the first row), to 3.0 (1.2)~$\mu\mathrm{G}$ at $t_\mathrm{look}=1.85$~Gyr (as shown in the second row). In addition, the satellite's magnetic field strength at apocentre (depicted in the second row) is also higher than the one in the isolated dwarf of 0.8 (0.5)~$\mu\mathrm{G}$ within 5 (10)~kpc (which will be further discussed in the next section).
Furthermore, the metallicity maps suggest that the ISM of the satellite does not mix significantly with the CGM of the host galaxy during infall, retaining most of its low-metallicity gas (i.e.\ lower than the metallicity of the CGM, enriched by outflows from the central galaxy).

\begin{figure*}
    \centering
    \includegraphics{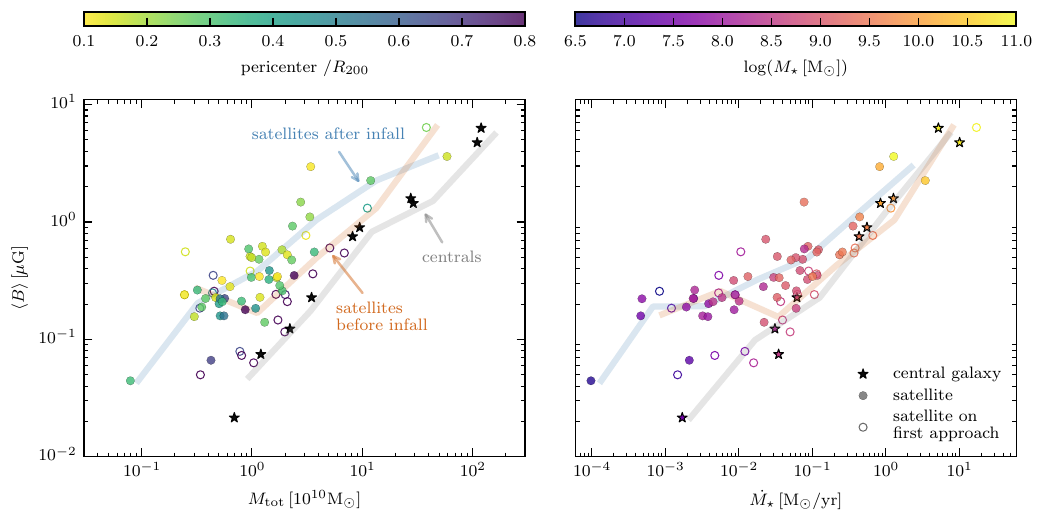}
    \caption{Average magnetic field strength (within a sphere with a radius of 20~kpc) of the satellite galaxies (circles) in comparison to central galaxies (star symbols) as a function of total mass (left-hand panel) and SFR (right-hand panel) at $z=0$ of the CR-MHD simulations. The solid lines represent the binned averages of the three types of galaxies shown here. We find systematically higher magnetic field strengths in the satellites compared to central galaxies at a fixed mass or SFR, except for the satellites that are still on their first approach to their host galaxy (open circles). The colours in the left-hand panel show the pericentres of the satellites in units of $R_{200}$ of their host halo, indicating a trend of larger magnetic field amplification for closer encounters with the host. All galaxies in the right-hand panel are colour-coded by their stellar mass, indicating a similar trend of stronger magnetic fields also as a function of stellar mass.}
    \label{fig:B_vs_Mass}
\end{figure*}

In Fig.~\ref{fig:B_vs_Mass}, we compare the averaged magnetic field strength of central galaxies to satellite galaxies at $z=0$ for the CR-MHD simulations, which we obtain by calculating the volume-weighted averaged magnetic energy density $\varepsilon_B=B^2/(8\pi)$ within a fixed radius of 20~kpc around the centre of all galaxies.\footnote{We note that our qualitative result does not depend on the exact choice of this radius. Moreover, the conclusions remain robust against choosing a mass-dependent radius (such as $0.5\times R_{200}$ or $3\times$ the half-mass radius).} Within the same region as the magnetic field, we calculate their SFR averaged over the last 100~Myr.
We only show centrals in the same stellar mass range as the satellites, i.e. $M_\star \lesssim 10^{11}\,\mathrm{M_\odot}$.
We discriminate between satellites on first approach and after that (i.e.~at least 300~Myr after the first pericentre of its orbit).

We find that satellites have systematically higher magnetic fields at $z=0$ in comparison to central galaxies with the same total mass and SFR, respectively. This effect is particularly pronounced if they have already had a close encounter with their host (see the binned averages shown as solid lines in Fig.~\ref{fig:B_vs_Mass}). In four mass bins from $M_\mathrm{tot}=10^{10}$ to $5\times10^{11}\,\mathrm{M_\odot}$ (equally spaced in log-space), the magnetic fields of satellites (after first infall) are enhanced on average by factors of 7.9, 6.0, 2.7, and 2.4, respectively, with a large scatter around those values. Similarly, in four SFR bins from $2\times10^{-3}$ to $9\times10^{-1}\,\mathrm{M_\odot\,yr^{-1}}$, magnetic fields are stronger on average by factors of 8.7, 3.2, 2.2, and 1.8.
We note that all central galaxies with total masses $M_\mathrm{tot}\leq10^{10.91}\,\mathrm{M_\odot}$ ($M_\star\leq 10^{9.48}\,\mathrm{M_\odot}$) are isolated dwarfs (see Table~\ref{table1}).
A similar trend is found when we compare the magnetic field strength as a function of stellar mass, which is indicated by the colour in the right-hand panel of Fig.~\ref{fig:B_vs_Mass}.
The encounter with the host seems to be crucial for this finding, since we do not obtain significantly enhanced magnetic fields for the satellites that are still infalling for the first time into the potential of the host and are still far away from the first pericentre of their orbit (i.e. with a distance to the centre of $D\gtrsim 0.7\,R_{200}$). 
In addition, we find a trend of enhanced amplification of the magnetic field with smaller pericentres of their orbit, i.e. how close the satellite approached the host galaxy, as indicated by the colour (quantifying the pericentre in units of $R_{200}$ of the host halo) in the left-hand panel of Fig.~\ref{fig:B_vs_Mass}. Here, the pericentre is defined as the minimum distance of the satellite to the central galaxy amongst all produced snapshots. If the satellite is still infalling, the colour corresponds to the current distance at $z=0$ in units of $R_{200}$.
For this subsample of galaxies, we likewise see a trend of enhanced amplification with decreasing distance to the host. 
Furthermore, we note that, during the pericentric passage, the magnetic field is temporarily increased at all radii, including the outskirts of the satellites. This is possibly because the sphere used for calculating the average could partially overlap with the magnetized CGM of the host. However, after pericentre, when the satellite's distance to the host is much larger than the considered averaging radius, the magnetic field is predominantly enhanced in the central region (see Fig.~\ref{fig:radial_profile}).
Fig.~\ref{fig:B_vs_Mass} depicts the results for our fiducial CR-MHD simulations with $\zeta_\mathrm{SN}=0.10$ and $\kappa=10^{28}\,\mathrm{cm^2\,s^{-1}}$, but we find the same trends for the simulations without CRs, as well as for variants of our CR runs with a smaller injection efficiency and larger diffusion coefficients. Hence, the result seems to be universal across our tested physical models.

\section{Physical processes setting the magnetic field }\label{sec:B-field-evolution-turbulent-driving}

\begin{figure*}
    \centering
    \includegraphics{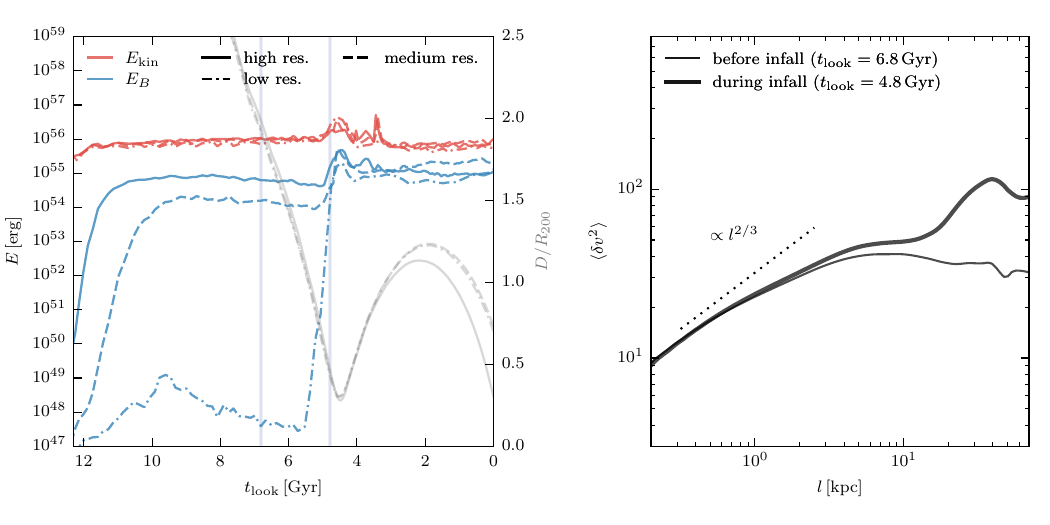}
    \caption{\textit{Left-hand panel}: Time evolution of kinetic and magnetic energies of the highest stellar mass satellite of halo 1e12-h12 at different resolution levels (`high res': $m_\mathrm{gas}=6\times10^3$; `medium res': $m_\mathrm{gas}=5\times10^4$; `low res': $m_\mathrm{gas}=4\times10^5$). The energies are integrated within a sphere with a radius of 20~kpc around the centre of the satellite. The grey lines show the distance $D$ to the central galaxy in units of $R_{200}(t)$ (right-hand $y$-axis). During the first infall, starting at $t_\mathrm{look}\sim5$~Gyr, the magnetic energy significantly increases for all resolutions and converges towards similar values at late times, where it remains enhanced compared to before infall. The light blue vertical lines on top of the $x$-axis indicate the times of the velocity structure functions shown in the right-hand panel. \textit{Right-hand panel}: Second order velocity structure function (see Eq.~\ref{eq:SF2}) of the high-res simulation before and during the infall of the same satellite, indicating a larger turbulent driving scale during infall than before infall. The theoretical scaling for Kolmogorov turbulence \citep{1941aKolmogorov} is indicated by the dotted line.}
    \label{fig:vel_structure_function_Energy_evolution}
\end{figure*}

To exemplify the temporal evolution of a typical satellite in our simulation, we show in Fig.~\ref{fig:vel_structure_function_Energy_evolution} the kinetic and magnetic energy of the satellite with the largest stellar mass of halo 1e12-h12 in our fiducial simulations (i.e.~the same satellite shown in Fig.~\ref{fig:map}) as a function of lookback time $t_\mathrm{look}$. 
The time evolution of both energies is shown for all resolution levels of this halo, together with the distance to the host halo (grey lines).
Before infall, the dynamo of the satellite is probably only driven due to processes like outflows and accretion, and strongly depends on resolution with orders of magnitude smaller magnetic energies in the lower resolution runs. 
However, during infall, i.e.\ close to the first pericentre of the orbit, we observe a significant increase in the magnetic and also kinetic energies, independent of the resolution, and converging towards comparable values of magnetic energy at $z=0$. Subsequently, we observe transient peaks in magnetic and kinetic energies corresponding to mergers and close encounters with other satellites.
The magnetic field not only gets amplified during infall, but remains higher compared to its strength before the satellite's first infall. This is consistent with the finding of higher magnetic fields for all satellites after infall compared to the satellites on their first approach (see Fig.~\ref{fig:B_vs_Mass}).

We observe a similar behaviour for the other three star-forming satellites of the high-resolution simulation of halo 1e12-h12. Their magnetic energies (within 20~kpc around their centres) are amplified by factors ranging from 1.3 to 5.4 when compared 1~Gyr before and after the first pericentric passage of their orbit.

The resolution-independent enhancement of the magnetic field can be understood by assessing the requirements for resolving a turbulent small-scale dynamo in numerical simulations. In a quasi-Lagrangian code, the outer driving scale of turbulence $\mathscr{L}$ relates to the numerical Reynolds number through \citep{2022Pfrommer}
    \begin{align}
        \mathrm{Re_{num}}\sim \frac{3 \mathscr{L} }{d_\mathrm{cell}},
        \label{eq:Re_num}
    \end{align}
where $d_\mathrm{cell}$ is the diameter of the Voronoi cell. This is derived for a quasi-Lagrangian code and assumes that the velocity scale at the turbulent injection scale is given by the thermal velocity \citep[see ][for details]{2022Pfrommer}.
To estimate the turbulent driving scale during the infall of the satellite, we calculate the longitudinal velocity structure function of second order $\langle \delta \varv^2 \rangle$ via 
\begin{align}
    \langle \delta \varv^2 \rangle (l) = \langle [\varv_{\parallel}(\bs{x} + \bs{l},t) - \varv_{\parallel}(\bs{x},t)]^2 \rangle,
    \label{eq:SF2}
\end{align}
where ${\varv}_{\parallel}$ is the velocity component along the separation vector $\bs{l}$ of two cells. We average over $\delta \varv^2$ in 100 log-spaced separation bins ranging from 0.2 to 100~kpc and calculate the parallel velocity difference for all pairs of gas cells within a sphere with a radius of 40~kpc around the centre of the satellite.\footnote{For this calculation, we do not take the subhalo membership of gas cells into account because in particular during the infall the subfind algorithm cannot reliably discriminate between bound gas cells of the satellite and the host galaxy.} 
We additionally calculate the transverse velocity structure function by using the velocity components perpendicular to $\bs{l}$, which exhibits qualitatively the same behaviour as the longitudinal one. Furthermore, we note that we obtain the same qualitative result when selecting cells within a larger radius of 50~kpc instead.

We show in the right-hand panel of Fig.~\ref{fig:vel_structure_function_Energy_evolution} the second-order velocity structure function for the same satellite as shown by the solid curves in the left-hand panel. While the structure function initially exhibits a flattening at around 5~kpc before its first infall, it shows a clear peak during the infall at $\sim 30$~kpc, indicating additional turbulent driving at this scale induced by the interaction of the satellite with the (turbulent) CGM of its host. The velocity structure functions are shown for the highest resolution level (i.e.\ level~3) of this halo, but they exhibit similar shapes when calculated for the lower resolution runs. 

To estimate the corresponding numerical Reynolds numbers from Eq.~(\ref{eq:Re_num}), we calculate the mean of the 10 (100) smallest cells of the satellite shown in Fig.~\ref{fig:vel_structure_function_Energy_evolution} and find for the lowest resolution simulation $d_\mathrm{cell}\approx$~439 (507)~pc, and for the highest resolution on average $d_\mathrm{cell}\approx$~90 (97)~pc. As a result, adopting $\mathscr{L}\sim5$~kpc before and 30~kpc during infall, we obtain for the highest resolution before and during infall $\mathrm{Re}_\mathrm{num}\sim160$ and  $\mathrm{Re}_\mathrm{num}\sim10^3$, respectively. However, the lowest resolution level exhibits $\mathrm{Re}_\mathrm{num}\sim30$ before infall, but during infall reaches $\mathrm{Re}_\mathrm{num}\sim180$.
Considering that we require numerical Reynolds numbers $\gtrsim 100$ \citep{2012Schober}, this naturally explains why the low-resolution simulations cannot resolve the small-scale dynamo before infall, while we obtain amplified fields for all resolution levels after the infall, because the driving scale is much larger at this time, and we reach $\mathrm{Re}_\mathrm{num}>100$ in all simulated cases. 

Hence, we conclude that while the turbulence is driven before infall on rather small scales, the additional turbulent driving at larger scales that satellites experience during a close encounter facilitates resolving a small-scale dynamo already at much lower resolution.
We therefore find converged magnetic fields within our satellites in the simulations with a gas mass resolution of $4\times10^5\,\msun$ (level~5) after first infall, while we require a resolution of $8\times10^2\msun$ (level~2) in isolated dwarfs in the simulations without CRs \citep{2024Pakmor}. Interestingly, we find faster convergence in dwarf galaxies in our simulations including CRs than without CRs (see Fig.~\ref{fig:B_resolution}), potentially due to some additional driving of outflows from CRs (Bieri et al.\ in preparation).

\section{Discussion}
\label{sec:discussion}

\begin{figure}
    \centering
    \includegraphics{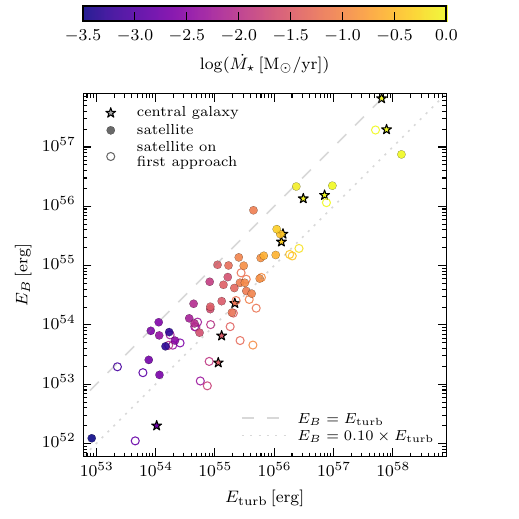}
    \caption{Total magnetic energy as a function of turbulent kinetic energy for the central galaxies (star symbols) and satellites (circles), colour-coded by their SFR. The dashed grey line shows equal energies, while the dotted grey line shows where $E_B$ is 10 per cent of $E_\mathrm{turb}$. While the magnetic energies of the highly star-forming central galaxies saturate at $\sim 20$ per cent of their turbulent energy, the isolated dwarfs only reach a few per cent. In contrast, the satellite galaxies yield much higher ratios of magnetic to kinetic energies than the isolated dwarfs at the same SFR. 
    }
    \label{fig:E_B_vs_E_kin}
\end{figure}

In a turbulent magnetic dynamo, we would expect the magnetic energy to saturate at several 10 per cent of the turbulent energy \citep[e.g.][]{2022Kriel, 2025Kriel}. Given more turbulent energy, one would expect higher magnetic fields.
Indeed, we find higher turbulent energies in satellites already before their first pericentric passage of their orbit compared to isolated dwarfs, with the same total mass, stellar mass or SFR (at $z=0$). 
This is likely caused by a more bursty star formation history in satellites than in isolated dwarfs, e.g.\ driven by interactions with other galaxies. 
This is reminiscent of pre-processing of galaxies infalling into galaxy clusters, such as the enhanced star formation activity that has been observed in galaxies on the outskirts of galaxy clusters \citep[e.g.][]{2012Mahajan, 2022Boselli, 2022Roberts, 2024Vulcani}. Consequently, some of our satellites already fall in with a slightly stronger magnetic field than the isolated dwarfs. Note that the magnetic field of the satellites simulated at level~4 resolution ($m_\mathrm{gas}=5\times10^4\msun$) is likely not converged before infall, and therefore represents a lower limit.
In particular, during the time of a close encounter satellites gain more kinetic energy (see e.g.\ the peak in kinetic energy in Fig.~\ref{fig:vel_structure_function_Energy_evolution}), which can potentially also induce more internal turbulent energy in the satellite.
Therefore, if they exhibit a larger reservoir of turbulent energy, and given sufficient numerical resolution, we would expect stronger magnetic fields compared to dwarfs with lower reservoirs of turbulent energy.\footnote{We caution at this point that our central dwarf galaxies are chosen to not have any nearby, more massive companion at $z=0$; hence, they probably will tend to have a more quiet merger history by construction in comparison to satellite galaxies.}

However, we find in most cases higher saturation strengths relative to the turbulent energy reservoir in satellites than in their isolated counterparts.
In Fig.~\ref{fig:E_B_vs_E_kin}, we show the total magnetic energy for our central and satellite galaxies at $z=0$, colour coded by their SFRs, as a function of turbulent energy. The latter is estimated as $E_\mathrm{turb}\sim 3/2 \times (E_\mathrm{kin,R} + E_\mathrm{kin,z})$, where $E_\mathrm{kin,R/z}$ are the kinetic energy in the radial and vertical direction, respectively. Here, the frame of reference is chosen such that the $z$-axis is aligned with the angular momentum vector of the stars. Furthermore, we assume that the kinetic energies in the radial and vertical directions are dominated by turbulence, and that the turbulent component in the azimuthal component is of similar size. The energies are integrated over all cells that are part of the corresponding subhalo, at $z=0$.
 
The central galaxies with $E_\mathrm{turb}>2\times10^{55}$~erg, with $M_\star>5\times10^8\,\msun$ and $\dot{M}_\star>0.06\,\msun/\mathrm{yr}$, reach magnetic energies that are at least 10 per cent of the turbulent energy. 
The central dwarfs with $M_\star < 5\times 10^8\,\mathrm{M_\odot}$ fall below $E_B/E_\mathrm{turb}=0.10$.
A few of the satellites that are on their first approach also have low magnetic energies (less than $10$ per cent of $E_\mathrm{turb}$). This is partly due to the fact that the resolution of the simulated haloes that the satellites belong to is not sufficient to resolve the small-scale dynamo driven on $\sim$~kpc scales before infall (as discussed in the previous section).
However, all satellites with a close encounter with their host in their past fall in the same range of $E_B/E_\mathrm{turb}$ as the more massive centrals. This suggests that they reach saturated magnetic fields, and that the interaction of a satellite with the host is necessary to make the dynamo saturate at the expected field strengths, at least within the Auriga feedback model.
Some of the satellites even reach a ratio of $E_B\approx E_\mathrm{turb}$.
Therefore, in addition to higher reservoirs of turbulent energy, additional mechanism(s) could be at play.

One potential contribution to higher field strengths in satellites could be mixing of their gas with the (magnetized) CGM of their host during close encounters. To quantify this, we follow the tracer particles in the satellites \citep{2013Genel}.\footnote{At the start of the simulation at $z=127$, there is one tracer particle per gas cell in the simulations. At $z=0$, typically, around 68 per cent of the gas cells in satellites have tracers.} Comparing the tracers 1~Gyr before and after the closest encounter with the central galaxy, we find that the fraction of cells (with tracers) that are part of the satellite after its infall but which have belonged to the central galaxy before infall lies between 4 and 23 per cent. This fraction of mixed cells neither correlates with the strength of the magnetic field of the satellite nor with the ratio of its pericentre to $R_{200}$. 
Furthermore, we find that the tracers of cells that are bound to the same subhalo before and after infall contain at least 79 per cent of the total magnetic energy of the satellite after infall.
This suggests that mixing is most likely not the main driver of the amplification but only partly contributes to it. This is also in agreement with the distribution of low-metallicity gas around the satellite in Fig.~\ref{fig:map}.

If the magnetic field is amplified due to adiabatic compression, we would expect a scaling of $B\propto \rho^{2/3}$. 
However, if the turbulent driving happens first in the outskirts of the satellite, where densities are low, we cannot disentangle this process easily from other processes happening at the same time.
Further potential contributions to the magnetic field amplification could include the exact timing and direction of infall, as well as the number of close encounters with the host. However, we find that the latter does not correlate with the magnitude of the amplified magnetic field strengths after infall. 
In particular, we do not find any significant additional amplification for the cases with multiple close encounters that goes beyond the amplification of the first close encounter.
Furthermore, we suggest that magnetic draping could potentially amplify the magnetic field in front of an infalling satellite, which might then partly mix with the satellite's ISM and contribute to higher field strengths there (see Fig.~\ref{fig:map}). While the mixing itself seems to be subdominant in our simulations, this effect and the respective importance compared to other mechanisms studied here should be investigated in future work. 

Irrespective of the driving mechanisms at play and their relative importance, stronger magnetic fields in satellite galaxies, or even more generally in interacting galaxies, would have major implications for the interpretation of observables related to CRs in star-forming galaxies. 
On the one hand, the strength of the magnetic field can strongly impact radio synchrotron emission. In particular, as long as $B < B_\mathrm{CMB}$, where the equivalent magnetic field of the cosmic microwave background (CMB) with an energy density $\varepsilon_\mathrm{CMB}$ is $B_\mathrm{CMB}= \sqrt{8\pi \varepsilon_\mathrm{CMB}} \approx 3\mu\mathrm{G}$, the synchrotron emissivity scales as $j_\nu \propto B^{2.05}$, which is fulfilled for most of our dwarfs (see Fig.~\ref{fig:B_vs_Mass}).\footnote{The scaling of $j_\nu$ with $B$ is derived from assuming an injected spectral index for CR electrons of 2.1 together with the fully-cooled limit where synchrotron losses dominate over other processes \citep{2008Pfrommer}.} It might therefore also be relevant for studying the conditions of electron calorimetry in dwarf galaxies, i.e.\ whether radiative losses dominate over escape losses \citep{2006Thompson,2010Lacki}.
On the other hand, due to the strong dependence of CR streaming and the related Alfv\'en losses for CRs on the magnetic field strength, it might also impact the calorimetry of CR protons. In turn, this would affect the resulting gamma-ray emission, and in particular also the conclusions drawn from comparing different CR transport models to observed gamma-ray luminosities \citep[][Werhahn et al.\ in preparation]{2019Chan, 2021WerhahnIII, 2022Nunez-Castineyra}.

\section{Conclusion}
\label{sec:conclusion}

In this work, we analyse the magnetic fields of satellite galaxies in cosmological zoom simulations. Our main conclusions are the following:
\begin{itemize}
    \item For satellite galaxies that have had a close encounter with their host, we find stronger magnetic field strengths in comparison to central galaxies, which in the mass range of the satellites are mainly isolated dwarfs. The stronger magnetic fields are both found as a function of total mass and SFR, respectively.
    We find this both in our simulations including CRs and without CRs, as well as in the simulations varying the CR transport and injection parameters.
    \item The interaction of the satellite with the CGM of the host seems to be crucial for this process because the enhancement of the magnetic field correlates with the pericentre of the orbit, and it is not as pronounced in the satellites which are still on their first infall.
    \item In contrast to the small-scale dynamo driven by turbulence due to feedback and accretion, which strongly depends on numerical resolution (and is not resolved in our lower resolution simulations), we find that the amplification of the field occurs during first infall of the satellites into the halo independently of the tested resolution.
    \item The velocity structure function points towards an additional turbulent driving at larger scales during the infall compared to before infall, which could explain the resolution independent amplification. This is because large numerical Reynolds numbers can already be reached at lower resolution if the driving scale is larger.
    \item Besides a generally more turbulent ISM in the satellites in comparison to dwarfs with the same mass (e.g.~because of interactions with other galaxies), further possible contributions to the additional amplification are adiabatic compression and/or magnetic draping, while we find that mixing with the magnetized CGM of the host is subdominant. Further disentangling these processes and quantifying their relative importance should be studied in future work in more detail.
\end{itemize}

Our results have important implications for other studies of satellite galaxies in simulations of large cosmological boxes which typically lack the resolution to resolve the internal small-scale dynamo of low-mass galaxies driven by turbulence on smaller scales. For those satellites that are past their first infall, the magnetic fields might already be converged at much lower resolution than previously assumed from studies of isolated dwarfs \citep[see][]{2024Pakmor}. Indeed, our results are in qualitative agreement with a study based on the TNG-100 simulation \citep{2025McDonough}.

Another important consequence of this result is the effect of higher magnetic field strengths in interacting galaxies on the resulting radio synchrotron emission. This might improve agreement with the observed far-infrared-radio correlation, particularly at the low-mass end, compared to previous isolated setups \citep{2021WerhahnIII, 2022Pfrommer}. 
Furthermore, due to the dependence of Alfv\'en wave losses on the magnetic field strength, stronger magnetic fields in interacting galaxies might also affect CR proton calorimetry and the related predictions of hadronic gamma-ray emission, which are often used to constrain CR transport parameters.
In particular, our results imply that using simulations of isolated (cosmological or non-cosmological) dwarf galaxies to compare to observables from satellites might strongly bias the conclusions. 
We aim to study both aspects, i.e.\ radio and gamma-ray observables, in follow-up works (Werhahn et al. in preparation).

\section*{Acknowledgements}
We thank Kandaswamy Subramanian for useful discussions about the velocity structure function. Furthermore, we thank the anonymous referee for a constructive report that helped to improve the paper.
RB is supported by the UZH Postdoc Grant (grant no. FK-23116) and the SNSF through the Ambizione Grant PZ00P2\_223532.
FvdV is supported by a Royal Society University Research Fellowship (URF\textbackslash R1\textbackslash191703).

\section*{Data Availability}
The data underlying this article will be shared on reasonable request to the corresponding author.


\bibliographystyle{mnras}
\bibliography{literature} 



\appendix

\section{Convergence study of magnetic fields}
\label{app:convergence-bfield}

\begin{figure*}
    \centering
    \includegraphics[]{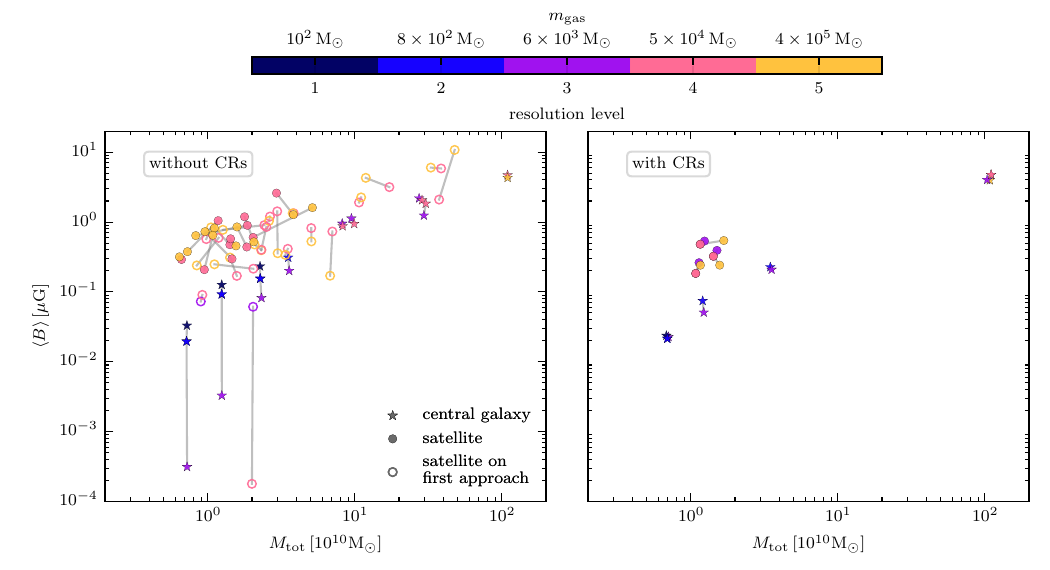}
    \caption{Magnetic field strength for different resolution levels as a function of total mass for satellites and central galaxies. The resolution levels 3, 4 and 5 correspond to `high res' ($m_\mathrm{gas}=6\times10^3\,\mathrm{M_\odot}$), `medium res' ($m_\mathrm{gas}=5\times10^4\,\mathrm{M_\odot}$) and `low res' ($m_\mathrm{gas}=4\times10^5\,\mathrm{M_\odot}$) in Fig.~\ref{fig:vel_structure_function_Energy_evolution}, respectively. While without CRs (left-hand panel), $m_\mathrm{gas}=10^2\,\mathrm{M_\odot}$ (`level 1') is required for convergence for isolated dwarfs with total masses of order $10^{10}\,\mathrm{M_\odot}$, including CRs (right-hand panel) leads to convergence already at lower resolution.
    }
    \label{fig:B_resolution}
\end{figure*}

In Fig.~\ref{fig:B_resolution}, we show the averaged magnetic field strengths within 20~kpc around the centre of all satellites and centrals galaxies (within the same range of total masses) of which we have run several resolution levels.
The left-hand panel shows the convergence for our runs without CRs \citep[see also][]{2024Pakmor}, where a resolution of at least level~2 (i.e. $m_\mathrm{gas}=8\times 10^2\,\msun$) is required for dwarf galaxies with masses $M_{200}\lesssim 10^{10}\,\msun$. For lower resolution levels, the magnetic fields of isolated dwarfs of these masses are underestimated by $\sim2$ orders of magnitude. 

However, the satellite galaxies already converge towards similar values for resolution level~5 and level~4, provided they are past their first infall. The convergence is, however, even stronger for our runs including CRs (right-hand panel of Fig.~\ref{fig:B_resolution}). Here, the obtained magnetic field strengths are close to identical in all centrals and satellites across all tested resolution levels. In particular, there is no significant change in the magnetic field strengths of the isolated dwarf galaxies when increasing the resolution from level~3 to level~2, and similarly, there is no change in the results for level~5 up to level~3 for the satellites.
Hence, we conclude that the results for the rest of our sample of satellites discussed in this paper, which we have only run up to resolution level~4, are robust for all satellites after first infall. 

\section{Power spectra}
\label{app:powerspectra}

\begin{figure}
    \centering
    \includegraphics[]{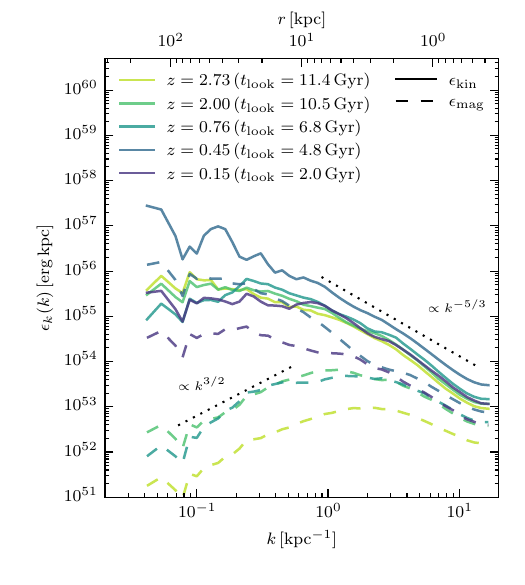}
    \caption{Magnetic (dashed lines) and kinetic (solid lines) energy power spectra at different times (as indicated in the legend) for the satellite galaxy analysed in Fig.~\ref{fig:vel_structure_function_Energy_evolution}. The magnetic power spectrum first grows at early times with a slope consistent with \citet{Kazantsev1985}, i.e.\ $\propto k^{3/2}$, until it saturates and exhibits a peak at $\sim5$~kpc at $z\sim2$. As the satellite gets close to the first pericentric passage (at $t_\mathrm{look}=4.8$~Gyr), the magnetic field grows both on small and on larger scales, where there is also additional power present in the kinetic energy spectrum. The latter exhibits a slope that is consistent with the scaling ($\propto k^{-5/3}$) as expected for subsonic turbulence \citep{1941aKolmogorov}.}
    \label{fig:powerspec}
\end{figure}

In Fig.~\ref{fig:powerspec}, we present the power spectra for the kinetic and magnetic energies for the satellite analysed in Fig.~\ref{fig:vel_structure_function_Energy_evolution}. We follow the same procedure as in \citet{2024Pakmor} for the calculation of the power spectra, except that we only choose a region with a radius of 40~kpc around the centre of the satellite, in order to be consistent with the analysis of the velocity structure function. 
At early times, before the satellite enters the virial radius of the host halo, the magnetic energy spectrum peaks at scales of around 5~kpc. During the time of pericentric passage of the satellite, we find increased magnetic power on all scales, but in particular on larger scales $\gtrsim20$~kpc.
Moreover, there is more energy present at larger scales during infall ($t_\mathrm{look}=4.8$~Gyr) in the kinetic energy spectrum, consistent with the peak in the velocity structure function at the same scales around 40~kpc, which we found in Fig.~\ref{fig:vel_structure_function_Energy_evolution} \citep[see also e.g.][]{2011Vazza,2017Vazza}.

\section{Radial profiles}
\label{app:radial_profiles}

\begin{figure}
    \centering
    \includegraphics[]{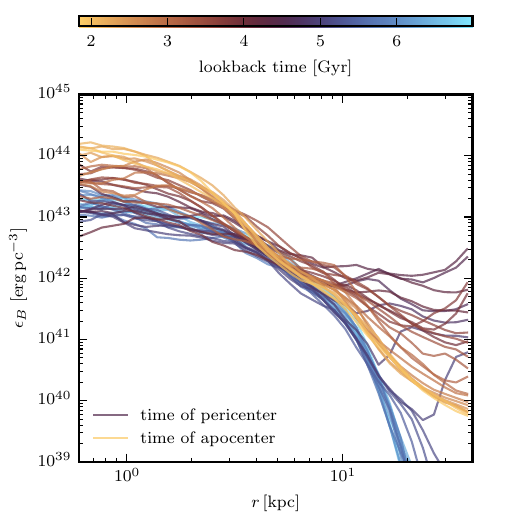}
    \caption{Radial profile of the magnetic energy density at different times (see colourbar). During the pericentric passage, the magnetic field increases in the outskirts, while there is a clear enhancement visible in the central region at later times.}
    \label{fig:radial_profile}
\end{figure}

The volume-weighted radial profiles of the magnetic energy density of the satellite analysed in Fig.~\ref{fig:vel_structure_function_Energy_evolution} and Fig.~\ref{fig:powerspec} are shown in Fig.~\ref{fig:radial_profile} at different times of evolution. 
The magnetic energy density temporarily increases in the outskirts during the first pericentric passage of the satellite around 5~Gyr ago, which is probably because of partial overlap with the CGM of the host galaxy. However, a clear increase of the magnetic energy density can be observed at subsequent times, in particular also in the central region ($r\lesssim5$~kpc) of the satellite.
This clearly shows that the increased magnetic field after infall, even if averaged over a sphere of 20~kpc, is dominated by the central region, which is amplified in comparison to before the first infall of the satellite.

\bsp	
\label{lastpage}
\end{document}